\documentclass[aps,prl,reprint,superscriptaddress]{revtex4-1}
\usepackage{graphicx}
\usepackage{siunitx}
\usepackage{natbib}

\begin{document}

\title{High Energy Electron Confinement in a Magnetic Cusp Configuration}

%

\author{Jaeyoung Park}
   \email[Corresponding author: ]{jypark@emc2fusion.com}
   \affiliation{Energy Matter Conversion Corporation (EMC2), 9155 Brown Deer Road, San Diego, CA 92121, USA}
\author{Nicholas A. Krall}
   \affiliation{1070 America Way, Del Mar, CA 92104, USA}
\author{Paul E. Sieck}
   \affiliation{Energy Matter Conversion Corporation (EMC2), 9155 Brown Deer Road, San Diego, CA 92121, USA}
\author{Dustin T. Offermann}
   \affiliation{Energy Matter Conversion Corporation (EMC2), 9155 Brown Deer Road, San Diego, CA 92121, USA}
\author{Michael Skillicorn}
   \affiliation{Energy Matter Conversion Corporation (EMC2), 9155 Brown Deer Road, San Diego, CA 92121, USA}
\author{Andrew Sanchez}
   \affiliation{Energy Matter Conversion Corporation (EMC2), 9155 Brown Deer Road, San Diego, CA 92121, USA}
\author{Kevin Davis}
   \affiliation{Energy Matter Conversion Corporation (EMC2), 9155 Brown Deer Road, San Diego, CA 92121, USA}
\author{Eric Alderson}
   \affiliation{Energy Matter Conversion Corporation (EMC2), 9155 Brown Deer Road, San Diego, CA 92121, USA}
\author{Giovanni Lapenta}
   \affiliation{Center for Mathematical Plasma Astrophysics, University of Leuven, Celestijnenlaan 200B, 3001 Leuven, Belgium}

\begin{abstract}
We report experimental results validating the concept that plasma confinement is
enhanced in a magnetic cusp configuration when $\beta$ (plasma pressure/magnetic field pressure) is
of order unity. This enhancement is required for a fusion power reactor based on cusp
confinement to be feasible. The magnetic cusp configuration possesses a critical advantage: the
plasma is stable to large scale perturbations. However, early work indicated that plasma loss
rates in a reactor based on a cusp configuration were too large for net power production. Grad
and others theorized that at high $\beta$ a sharp boundary would form between the plasma and the
magnetic field, leading to substantially smaller loss rates. The current experiment validates this
theoretical conjecture for the first time and represents critical progress toward the Polywell
fusion concept which combines a high $\beta$ cusp configuration with an electrostatic fusion for a
compact, economical, power-producing nuclear fusion reactor.
\end{abstract}

\maketitle 

\section{Background}
The use of magnetic fields to confine high temperature plasmas has been one of the main
pathways pursued in controlled thermonuclear fusion research since the 1950s. Several magnetic
field configurations such as magnetic pinch, stellarator, magnetic mirror, and tokamak, have
been explored to achieve net power generation from fusion reactions\cite{Bishop1958,Tuck1960,Wesson2011}. However, one of the
critical technical challenges related to magnetically confined fusion devices is the plasma
instability inside the confining magnetic fields. For example, magnetohydrodynamic (MHD)
instabilities driven by plasma current or plasma pressure such as kink and Rayleigh-Taylor
instabilities can abruptly disrupt the plasma confinement by tearing apart confining magnetic
fields and expelling the plasma. Such plasma instabilities limit the maximum operating plasma
current or pressure in the device and increase the reactor size required to achieve net fusion
power. Moreover, a large engineering safety margin is typically required to prevent reactor
failure in the event of a major disruption, increasing engineering complexities and reactor cost.

In comparison, the magnetic cusp configuration provides excellent macroscopic plasma
stability due to the convex magnetic field curvature towards the confined plasma system in the
center, as shown in Figure 1A.\cite{Bishop1958,Tuck1960,Grad1955} Experiments on the cusp field configuration have confirmed
the stability property, even at very high plasma pressures up to $\beta=1$.\cite{Spalding1971,Haines1977} Plasma beta, $\beta$, is defined
as the ratio of plasma pressure to confining magnetic field pressure, $\beta=P_{plasma}/(B^2/2\mu_0)$, where
$P_{plasma}$ is the plasma pressure, $\mu_0$ is the magnetic permeability, and $B$ is the magnetic flux density.
Since the fusion power output scales as $\beta^2$ for a given magnetic field, high $\beta$ operation is
advantageous for a compact economical fusion reactor. In comparison, the design parameter for
the International Thermonuclear Experimental Reactor (ITER), a proposed tokamak device to
achieve a net fusion power output, is $\beta \approx 0.03$.\cite{ITER1999}

\begin{figure}
   \includegraphics[trim=0 15 0 0,clip,width=\linewidth]{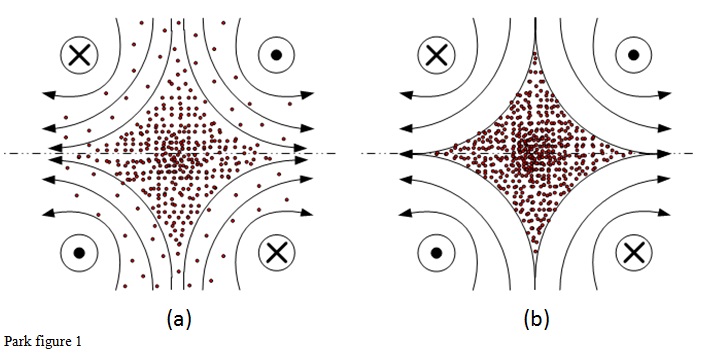}
   \caption{Comparison of low $\beta$ and high $\beta$ plasma confinement in a magnetic cusp configuration: (a) low $\beta$; (b) high $\beta$.}
\end{figure}

Substantial theoretical and experimental efforts were devoted to investigating the
magnetic cusp configuration.\cite{Bishop1958,Wesson2011,Grad1955} Initial results, however, showed poor plasma confinement.\cite{Bishop1958}
This was thought to be related to the open magnetic field structure and rapid mirror-like plasma
loss in a low $\beta$ cusp. Grad and others predicted theoretically that the plasma confinement
properties of the cusp configuration would be greatly enhanced if the magnetic field exhibits a
sharp boundary separating the field-free high $\beta$ plasmas and the vacuum region with magnetic
fields, as shown in Figure 1B.\cite{Bishop1958,Berkowitz1958} Equation 1 describes the theoretically estimated plasma loss
rate for the cusp system in Figure 1B.\cite{Berkowitz1958} The physical idea behind Equation 1 is as follows:
When $\beta$ is large there is a sharp transition between the confined plasma and the confining
magnetic field. Plasma approaching this transition layer reflects back into the confined volume.
Eventually, however, a plasma particle after many reflections will move almost exactly in the
direction of the cusp opening, and will be lost. Grad conjectured that this loss hole will have a
radius equal to the electron gyro-radius, as shown in Equation 1. By contrast, when $\beta$ is small in
the cusp, the transition region is the size of the confined volume, and plasma approaching the
boundary attaches to field lines, and streams out the cusp. This loss rate is related to the plasma
loss rate in a magnetic mirror and is much larger than the rate given in Equation 1.\cite{Krall1995}

Equation 1: Electron and ion loss rate for a single cusp during high $\beta$ plasma state
\begin{equation}
   \frac{I_{e,i}}{e} = \frac{\pi}9 n_{e,i} v_{e,i} \times \pi (r_{e,i}^{gyro})^2
\end{equation}
where $I$ is the loss current, $e$ is the electron charge, $n$ is the density, $v$ is the velocity, $r^{gyro}$
is the local gyro-radius at the cusp location, and subscript $e$ and $i$ denote electron and ion species
respectively.

Though several experiments were constructed to validate this conjecture, two critical
issues limited their efforts.\cite{Spalding1971,Haines1977,Marshall1960,Pechacek1980} The first issue was the engineering and technical challenge
related to initially forming a high $\beta$ plasma, where a required initial injection power is on the
order of 100 MW or more. The second issue was the theoretical and experimental difficulty in
determining the plasma loss rate in a high $\beta$ plasma state. It was accepted that if the loss rate is
determined by the ion gyro-radius, it would be unacceptably large for a fusion power reactor.
Experiments seemed to indicate that the ion gyro-radius did indeed dominate the loss rate.\cite{Pechacek1980}
Because of these problems, the concept of a fusion power reactor based on a cusp magnetic field
was largely abandoned, until a new idea, discussed in the next section, was proposed that retains
the advantages of cusp confinement, but removes the issue of the ion gyro-radius dominating the
loss rates.

\section{The Polywell Fusion Concept}
In 1985, Bussard introduced a fusion concept, the ``Polywell'' reactor, which combines
the magnetic cusp configuration with the inertial electrostatic confinement (IEC) fusion
concept.\cite{Bussard1991,Krall1992,Bussard2006} In the Polywell reactor, electrons are confined magnetically by a cusp field while
ions are confined by an electrostatic potential well produced by electron beam injection. The use
of an electron beam provides two critical advantages for the Polywell reactor over other
magnetic cusp devices. The excess electrons from the beam form an electrostatic potential
well.\cite{Farnsworth1968,Hirsch1967,Elmore1959} By utilizing an electrostatic potential well, the Polywell reactor employs a highly
efficient method to accelerate ions to high energies for fusion. In addition, the potential well
reduces ion kinetic energy as the ions travel outward toward the cusp exits. This results in
electrostatic confinement of ions as well as reduction in the ion loss rate given in Equation 1. In
a Polywell reactor, the main issues of high temperature plasma containment and plasma heating
are thus reduced to the confinement property of the injected electron beam. In work to date, a
deep potential well was formed and maintained in a cusp field.\cite{Krall1995} But that work was limited by
the high loss rates typical of a low beta cusp confined plasma. To progress further it is necessary
to show that electron confinement is enhanced at high $\beta$, as conjectured by Grad and others.

\section{Confinement Measurement at High $\beta$}
In this paper, we present the first experimental confirmation of the idea that confinement
of injected electron beam by a magnetic cusp is enhanced in a high $\beta$ plasma state. Figure 2
shows the experimental setup based on a hexahedral, or 6-coil, Polywell system. Two high
power pulsed co-axial plasma guns are used to inject plasma to form a high $\beta$ plasma in the
interior of the cusp configuration. An electron beam injector is based on lanthanum hexaboride
($\mathrm{LaB_6}$) thermionic emitter and produces 1--3~A of electron current at 7~keV beam energy. The
plasma guns produce the high $\beta$ condition and the electron beam measures the confinement
property.

\begin{figure*}
   \includegraphics[trim=0 25 0 0,clip,width=0.75\linewidth]{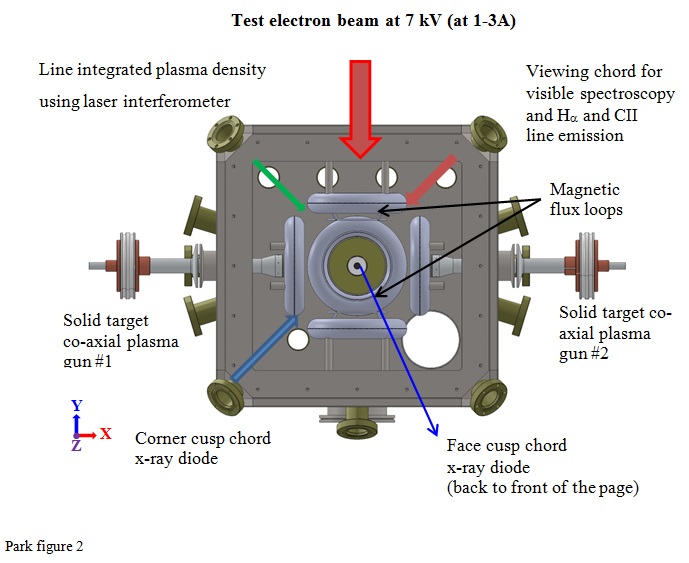}
   \caption{Experimental setup of high $\beta$ plasma confinement study used in the present article
   using hexahedral cusp configuration.}
\end{figure*}

The confinement of the injected electron beams is measured by x-ray diodes via electron
beam induced Bremsstrahlung. Improved confinement should lead to a higher density of high
energy beam electrons, which would give a stronger x-ray signal. Other plasma diagnostics
include a) line integrated laser interferometry for plasma density, b) two magnetic flux loops for
plasma diamagnetism, c) time resolved visible light emission spectroscopy for plasma ion
composition and plasma density, and d) two visible photodiodes for H$\alpha$ and C~II line emission
intensities. Further details are available in the Supplementary Methods section below.

\begin{figure*}[]
   \includegraphics[width=0.75\linewidth]{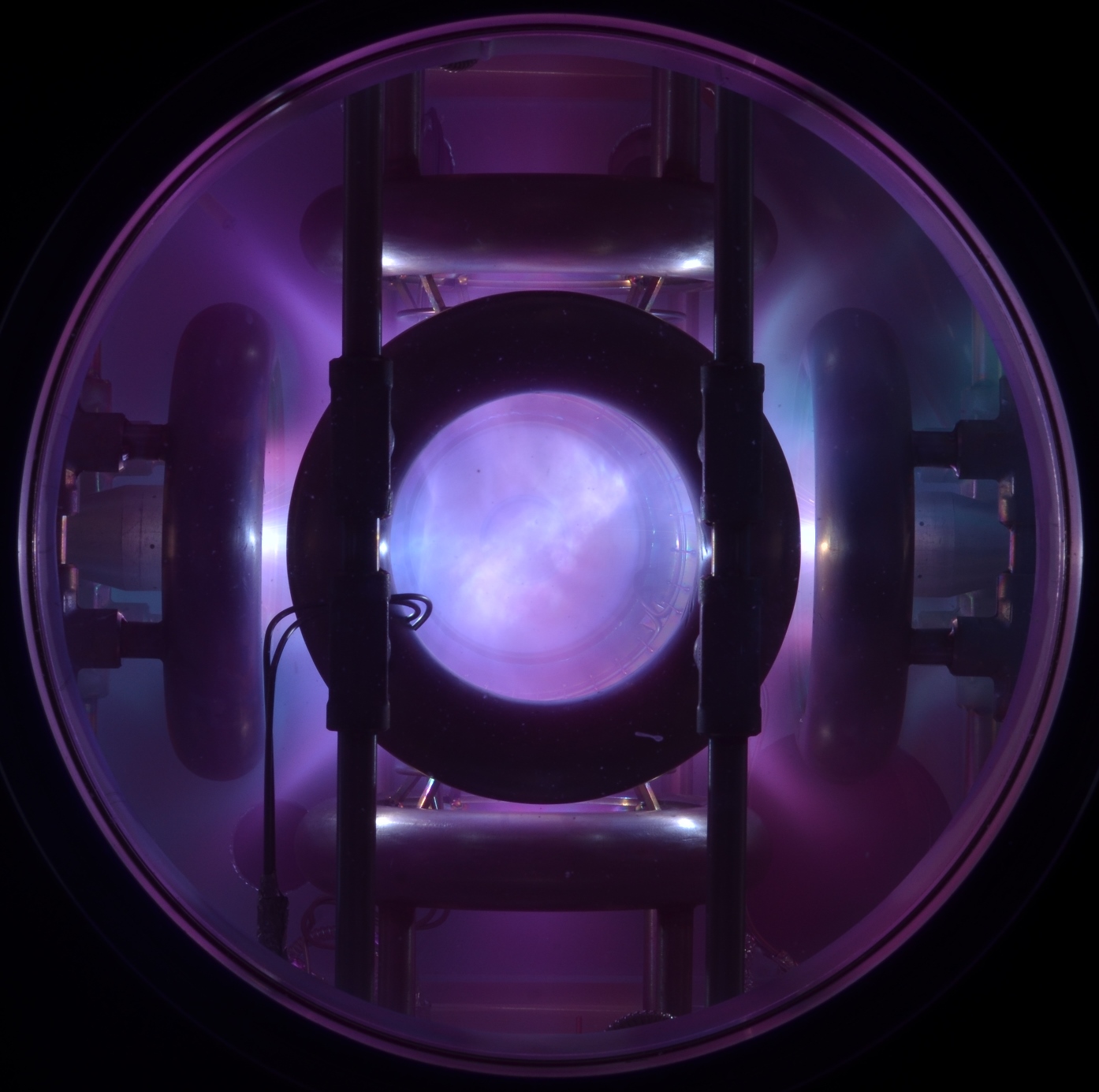}
   \caption{Time integrated raw visible light image of plasma from a high $\beta$ shot (\#15640).}
\end{figure*}

Figure 3 shows a time integrated visible image of a high $\beta$ plasma in the cusp system and
Figure 4 shows time resolved experimental results from the various diagnostics from a high $\beta$
shot, 15640. The coils are energized 40~ms prior to plasma injection and the coil currents are
kept at a constant value during the time period shown in Figure 4, with the B-field value of \SI{2.7}{kG}
at the cusp points near the plasma guns. In addition, the electron beam is turned on 30~$\mu$s
before gun plasma injection and operates with $\sim$3~A of injection current at 7.2~kV until t=150~$\mu$s.
At t=0, two co-axial plasma injectors are switched on with 700~MW combined peak input power
for $\sim$7~$\mu$s. Despite ringing in the underdamped gun circuit, the plasma injection provides high
density plasma in the cusp system with a rapid rise and a gradual decay as measured by a
heterodyne laser interferometer. The interferometer results were consistent with the plasma
density estimate by Stark broadening of the H$\alpha$ line. Plasma density, marked $\mathrm{n_e^{bulk}}$, increases to
\SI{9e15}{cm^{-3}} at t=9 $\mu$s as the plasma from the injectors is successfully transported to the center of
the magnetic cusp system. The plasma density stays nearly constant until t=20~$\mu$s and decreases
gradually to \SI{5e15}{cm^{-3}} at t$\approx$25~$\mu$s and \SI{2.5e15}{cm^{-3}} at t$\approx$32~$\mu$s. Flux exclusion signals from
flux loops, marked as $\Delta\Phi$, show a clear sign of an electron diamagnetic effect from the high $\beta$
plasma formation in the cusp magnetic fields. The flux loop data peaks at \SI{1.4e-4}{Wb} of flux
exclusion at t=12.7~$\mu$s and decreases to \SI{0.4e-4}{Wb} at t$\approx$21~$\mu$s. It is noted that \SI{1.4e-4}{Wb} of
excluded flux is equivalent to 10\% of the vacuum magnetic flux at the location of the flux loop.
Separately, visible photodiode data also indicates the injection of plasma and its gradual decay
based on H$\alpha$ and C II line emission intensity. The faster decrease in the C II line intensity
indicates cooling of the injected plasma after initial injection.

\begin{figure*}
   \includegraphics[trim=0 25 0 0,clip,width=0.7\linewidth]{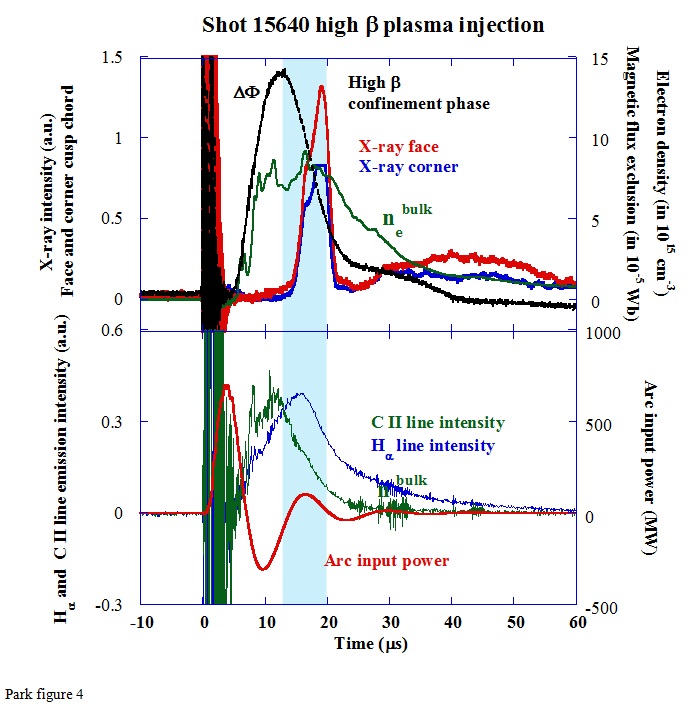}
   \caption{Time resolved experimental results showing enhanced high energy electron
   confinement of high $\beta$ plasma state in a magnetic cusp configuration.}
\end{figure*}

In comparison, the hard x-ray emission intensity shows very distinctive features in the
time domain. Prior to plasma injection, the x-ray diode signals between t=${-10~\mu s}$ and t=0
provide an estimate for the x-ray background data since there are no plasma ions to produce
Bremsstrahlung during this time period. The low background signals in the x-ray diodes
demonstrate good spatial collimation of x-ray detectors. Covering any metallic surfaces in the
line of sight of the x-ray detectors with plastic materials keeps K$\alpha$ emission from these surfaces
below the Kapton filter threshold energy of 2~keV. Initially, the x-ray signals remain low until
12~$\mu$s after plasma injection although the plasma density reaches its peak value of \SI{9e15}{cm^{-3}} at
t=9~$\mu$s. Shortly after the peaking of flux loop data at 12.7~$\mu$s, both x-ray diodes register strong
increases in hard x-ray emission, while the bulk plasma density varies little. The increase in xray
emission builds up from t=13~$\mu$s to t=19~$\mu$s, while all other plasma diagnostics indicate
gradual decay of injected plasma in the cusp. At t=19~$\mu$s, the x-ray emission intensity from both
viewing channels rapidly decreases toward zero within 1.5~$\mu$s. In contrast, the plasma density
shows only gradual decrease during that time period and there is no sudden change in any other
diagnostic. It is noted that the fast oscilloscope range was set too low for the x-ray signal
viewing through the cusp corners of the coils, resulting in artificial saturation in the x-ray signal.

This temporal behaviour of the x-ray emission signals can be explained as follows,
clearly demonstrating the relation between high $\beta$ plasma in the cusp magnetic field and the
improved electron beam confinement as postulated by Grad and others.\cite{Bishop1958,Grad1955,Berkowitz1958} Initially, the beam
electrons are confined poorly in the magnetic cusp system with a low $\beta$ plasma, resulting in low
beam density and low x-ray emission. After the intense plasma injection, the cusp magnetic field
is altered by the diamagnetism of the high $\beta$ plasma. This altered configuration, as in Figure 1B,
gives enhanced electron confinement, as predicted theoretically. The increase in hard x-ray
emission shows that beam electrons are now better confined in the magnetic cusp in the presence
of high $\beta$ plasma and the beam electron density starts increasing in time. A preliminary estimate
indicates that beam electrons are confined for at least 300 bounces in the cusp at high $\beta$,
compared to $\sim$7 bounces at low $\beta$. Since the beam power is limited at 20~kW, the increase in
beam density and therefore the x-ray emission intensity is gradual in time. The present
experimental setup does not have a subsequent plasma heating system after the initial plasma
injection, therefore the high $\beta$ plasma cannot be maintained indefinitely against plasma cooling
and eventual loss of injected electron density. The enhanced plasma confinement phase is only
temporary and it soon reverts back to the poor confinement phase when plasma $\beta$ becomes low.
When this transition occurs, all the previously confined high energy electrons will rapidly leave
the magnetic cusp, causing the observed rapid decrease in x-ray emission at t=21~$\mu$s. It is noted
that if the electron beam injection has sufficient power, it may prolong the duration of a good
plasma confinement regime.

The key advance of the current work in validating the high $\beta$ plasma cusp confinement
property stems from the use of an electron beam and beam induced hard x-ray measurement to
measure the confinement property. This is because an electron beam at 7~keV has a transit time
of 7~ns to move across the cusp system, compared to several microseconds during which the high
$\beta$ state is maintained. This provides the temporal resolution necessary to measure the change in
confinement property of the beam electrons by the magnetic cusp system during the plasma
injection. This feature was absent in all previous magnetic cusp experiments. In comparison,
high energy electron confinement is a critical issue in the Polywell fusion concept, which
facilitated the use of electron beam as a diagnostic to measure confinement. In addition, the
beam electrons have minimal coupling with the dense background plasma due to their high
energy and low density, which avoids diagnostic complexities in separating losses of ions and
electrons.

Figure 5 and 6 show variation in x-ray emission characteristic as plasma injection power
and cusp magnetic flux density are varied. The plots were generated using the x-ray signals from
viewing the cusp via the centers of coils; the characteristics of x-ray signals from the corners of
cusps are qualitatively identical. In Figure 5, the plasma injection power is varied from 220~MW
to 700~MW for a constant coil current generating 2.7~kG of magnetic field at the cusp points near
the plasma injectors. The x-ray signal results show that improvement in electron beam
confinement occurs only when there is a sufficient input power or energy in the plasma injection.
In comparison, the excluded flux and bulk plasma density increase with plasma injection power.
Figure 6 shows the x-ray emission as a function of cusp magnetic field for a constant plasma
injection power at 700~MW. At B=0, there is no diamagnetism and the x-ray signal shows no
distinctive structure, despite a clear indication of plasma injection from the density measurement
albeit at a low level. It is noted that weak x-ray emission with slow and gradual rises and decays
is explained by x-ray Bremsstrahlung from high Z impurities such as tungsten from the coaxial
gun electrodes. This has been verified by time resolved visible emission spectroscopy using a
gated CCD measuring various tungsten impurity line intensities compared to C line intensities.
At B=0.6~kG, the x-ray result shows changes in electron beam confinement property but the
signal is complex to interpret. In comparison, the x-ray result clearly shows a well-defined
period of large increase in electron beam confinement at B=2.7~kG. The excluded magnetic flux
shows faster peaking and bigger diamagnetic effects at B=0.6~kG compared to B=2.7~kG. The
peak flux exclusion for the B=0.6~kG case is equivalent to 44\% of the vacuum magnetic flux,
compared to 10\% for B=2.7~kG. The injected plasma density is comparable initially, while the
density decays rather rapidly for B=0.6~kG. Though the detailed analysis of these results are
beyond the scope of this work, the results indicate a level of balance between the magnetic field
pressure and the injected plasma pressure may be a key to an improvement in electron beam
confinement.

\begin{figure*}
   \includegraphics[trim=0 25 0 0,clip,width=0.7\linewidth]{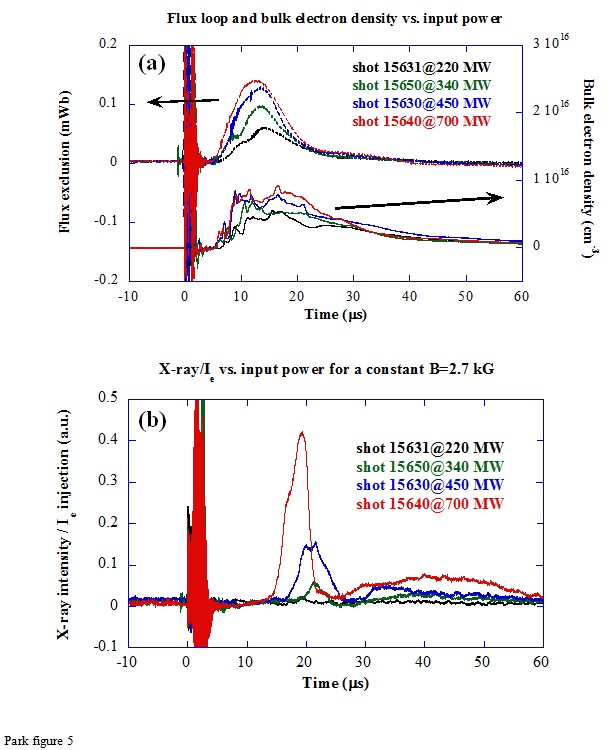}
   \caption{Plasma parameters as a function of plasma injection power at B-field of 2.7~kG: (a)~flux exclusion and bulk electron density, (b)~normalized x-ray emission intensity.}
\end{figure*}

\begin{figure*}
   \includegraphics[trim=0 25 0 0,clip,width=0.7\linewidth]{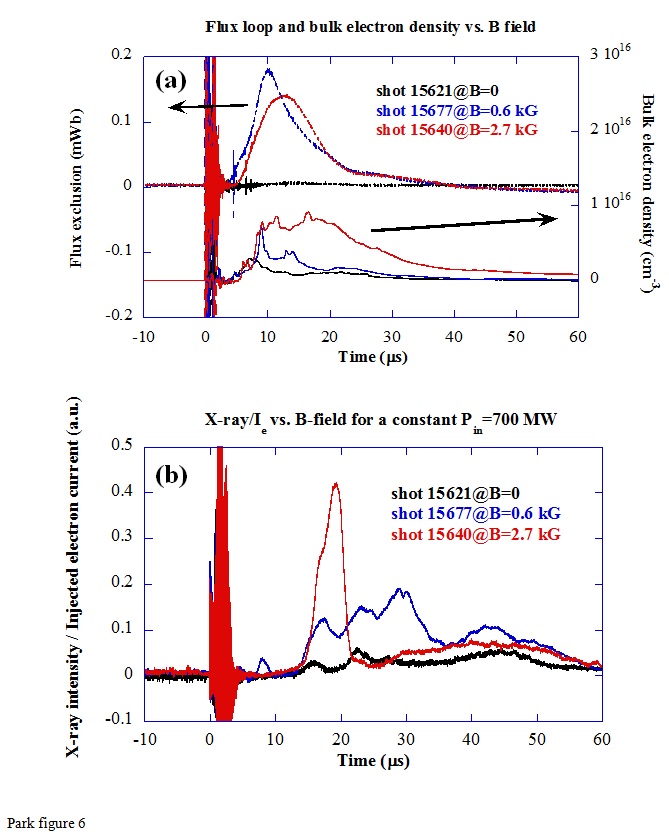}
   \caption{Plasma parameters as a function of magnetic field at 700~MW injection power: (a)~flux exclusion and bulk electron density, (b)~normalized x-ray emission intensity.}
\end{figure*}

\section{Computations and Scaling Implications}
The theory of high $\beta$ plasma injection and electron beam confinement has also been
investigated using the three dimensional particle-in-cell code, IPIC3D. IPIC3D is a massively
parallel code solving the Vlasov-Maxwell system of equations using an implicit time scheme and
has been used extensively in simulating magnetic reconnection and space plasma weather.\cite{Markidis2010}
Initial results from IPIC3D were promising in that high $\beta$ plasma injection resulted in strong
diamagnetic effects and a significant change in magnetic field in a cusp. At present, the
numerical study of confinement is limited by the spatial resolution available and the number of
particles required to reduce the noise level. As a result of these limitations, available
computational resources limited the range of initial $\beta$ vales where the improved confinement is
limited. Further discussion and results of these simulations are available with the Supplemental
Methods below.

The present experimental result is a major step toward a Polywell fusion reactor in that it
validates the conjecture that high energy electron confinement is improved in a high $\beta$ plasma.
However, two additional measurements are needed to estimate the performance of a Polywell
fusion reactor. The first is to quantitatively determine the loss rate. The second is to measure the
efficiency of ion acceleration by electron beam injection. For the purpose of discussion, we
estimate the power balance for a 1 meter radius hexahedral D-T Polywell fusion reactor
operating at $\beta$=1 with a magnetic field of 7~T at the cusp points and an electron beam injection
energy at 60~kV. This calculation is based on two assumptions: 1) the electron loss rate in
Equation 1 is correct, and 2) the efficiency of ion acceleration via a potential well can be made
good enough to convert 50\% of electron beam injection energy into an average ion energy.
From Equation 1, the electron loss current is 254~A per cusp for the electron density \SI{2e15}{cm^{-3}}
and electron energy at 60~keV at the cusp points. Since there are 14 cusps in the hexahedral
system, the required electron beam power to maintain a $\beta$=1 state would be 213~MW.
Separately, this system will lose an additional 51~MW of power via Bremsstrahlung radiation for
an average electron temperature of 60~keV, assuming no ions other than hydrogen isotopes are
present.\cite{Huba2013} In comparison, the expected D-T fusion power output would be 1.9~GW for a D-T
cross section of 1.38~barns at a center of mass energy of 30~keV.\cite{Bosch1992} Though speculative at this
point, this simple power balance scaling, coupled with the observed good plasma stability of a
magnetic cusp system, indicates that the Polywell may emerge as an attractive concept for a
compact and economical fusion reactor.

\section{Conclusions}
In summary, the present experimental results demonstrate for the first time that high $\beta$
plasma operation can dramatically improve high energy electron confinement in the magnetic
cusp system. This result validates the central premise of the Polywell fusion concept which uses
high energy beam injected electrons to create an electrostatic potential well for ion acceleration
and confinement. The current plan is to extend the present work with increased electron beam
power to sustain the high $\beta$ plasma state and to form an electrostatic well. If the deep potential
well can be formed and the scaling of the electron beam confinement is found to be favourable,
as conjectured by Grad and others, it may be possible to construct a compact, low cost, high $\beta$
fusion power reactor based on the Polywell concept.

\bibliography{Park_hi_beta_cusp}

\begin{acknowledgments}

We would like to thank John Santarius at U. of Wisconsin for his
theoretical support and discussions, Malcolm Fowler at McFarland Instrumentation for his
assistance in x-ray measurements, Greg Dale and Robert Aragonez at Los Alamos National
Laboratory (LANL) for their assistance in pulsed power system construction, Glen Wurden at
LANL for his assistance in design and operation of visible spectroscopy and x-ray diagnostics,
Kim Gunther and Marc Curtis at Heatwave and Ken Williams at Applied Science and
Engineering for their assistance in development of the electron beam injector system, Elizabeth
Merritt and Mark Gilmore at University of New Mexico for their assistance in development of
laser interferometer, and Mike Wray, Noli Casama and Kevin Wray at EMC2 for their laboratory
operation support.

This work was performed under Contract N68936-09-0125 awarded by the United States Department of Defense.

\end{acknowledgments}

\section{Author Contributions}
J.P designed experiments and performed data analysis, N.A.K. provided
theoretical support, and assisted in experimental planning and data analysis, P.E.S. performed
pulsed power system design and experimental operation, and provided magnetic flux loop
measurements, D.T.O. provided x-ray, visible spectroscopy and plasma density measurements,
M.S. provided mechanical design of the experiments, A.S. assisted in numerical simulation and
power system design, K.D. assisted in experimental operation and power system design, E.A.
assisted in mechanical design of the experiments, G.L. performed numerical simulation and
interpretation.

\section{Supplementary Methods}
\setcounter{figure}{0}
\renewcommand{\thefigure}{S\arabic{figure}}
\subsection{Magnetic cusp system}
The magnetic cusp experiments described in this paper were conducted in a cubic
vacuum chamber measuring 45~cm on an edge. Centred in this chamber are six identical magnet
coils, each coil having major radius 6.9~cm and minor radius 1.3~cm. The coils are arranged such
that each coil is centered on a face of a cube measuring 21.6~cm on an edge as shown in Figure 2,
producing cusp magnetic field. Each coil is driven by its own battery powered system capable of
generating a static magnetic field from 0.6~kG to 2.7~kG at the cusp near the coil center with coil
currents between 5--22~kA turns. The coils are energized for pulse duration of 40~ms compared
to the ~100 $\mu$s plasma duration, thus generating static magnetic field during the time of interest.

\subsection{Plasma injection system}
The plasma injectors consist of two co-axial plasma guns with an anode to cathode gap
spacing of 2~mm utilizing $\vec{j}\times\vec{B}$ plasma acceleration.\cite{Marshall1960} The plasma
guns use solid polypropylene film with 4~$\mu$m thickness instead of more common gas injection to generate
high pressure plasmas in a very short time. The amount of material injection is controlled by the number of
film layers; two layers were chosen for the work presented here. The plasma guns are
constructed with tungsten electrodes and boron nitride insulators to minimize erosion and
impurity injection. Each gun is powered by a high voltage capacitor and operates with 60--150~kA
of gun current using an Ignitron switch. Each gun is capable of injecting high pressure
plasma with up to 500~MW pulse power from the capacitor and pulse duration between 5--10~$\mu$s.
The plasma guns are located 0.5~cm outside the magnetic cusp points along the x-axis.

\subsection{Electron beam injector}
The electron beam injector utilizes a Lanthanum hexaboride ($\mathrm{LaB_6}$) cathode to produce
thermionic electron emission in the plasma environment. Beam extraction was obtained with the
use of a triple grid system made of graphite grids to provide controlled electron extraction, while
limiting plasma bombardment to the cathode surface. The electron beam is pulsed for 180~$\mu$s,
starting from 30~$\mu$s before the plasma injection. The electron beam injector is located 50~cm
above the top coil along the Y-axis.

\subsection{X-ray detection system}
The x-ray detection system consists of a biased photodiode in a differentially pumped
housing, a Kapton black window, a plastic collimator, and a magnetic yoke. The Kapton
window was 25~$\mu$m thick and completely opaque to visible light as well as x-rays below 2~keV.
The collimator was designed to maximize the core plasma volume in the line of sight for the
photodiode, while preventing the detection of x-rays produced when beam electrons are incident
on the coil surfaces. The magnetic yoke with 2~kG magnetic field is used near the entrance of
the collimator to prevent the beam electron induced fluorescence of the Kapton window. Two
viewing chords are used; one viewing the central confined plasma through the cusp openings in
the middle of coils along z-axis and the other viewing the plasma through the cusp openings
among the corner of coils along the x=y=z line. The former view has a collection volume of \SI{620}{cm^3}
and the latter view has a collection volume of \SI{380}{cm^3}.

\subsection{Laser interferometer}
A heterodyne laser interferometer was used to measure a line integrated plasma density
through the corner cusp using a green laser at 532~nm. The plasma length is assumed to be
22~cm, equal to the cusp system diameter. At that length, an average electron density of
\SI{1e16}{cm^{-3}} is equal to a phase shift of 180 degrees.

\subsection{Time resolved visible spectroscopy}
An Acton monochrometer with selectable gratings (2400, 600, and 300 grooves/mm)
was used in conjunction with a Princeton Instruments PI-Max intensified CCD camera to acquire
time resolved spectroscopic data for Stark broadening and plasma chemical composition.

\subsection{Magnetic flux loop}
Two magnetic flux loops are installed to monitor the plasma diamagnetism. Each loop
consists of a single circular turn of a coaxial cable with major radius 5.2~cm and minor radius
1.1~mm. The loops are located 1~cm away from the magnet casing surface, such that the plane of
each loop is 8.0~cm from the chamber center.

\subsection{Photodiodes for H$\alpha$ and C II line emission}
Two optical fibers collected plasma photoemission from the corner of the Polywell to
photodiode detectors. Narrowband interference filters were used on each detector, one for the H$\alpha$
line and the other for the C~II line at 724 nm.

\subsection{IPIC 3D numerical simulation}
Since the physics of diamagnetism is not related to the Debye length scale, we chose the
implicit method that relaxes the length and time scale requirement of the simulation, compared to
the explicit PIC model. The present simulation resolves the electron cyclotron time scale and the
electron skin depth, instead of the electron plasma frequency and the Debye length. The
simulation utilizes an electron mass of 1/256 of the hydrogen ion mass.
The simulation utilizes the same hexahedral geometry of the experiments with the
following changes. The total simulation size is 38~cm cube with a conducting metal boundary
corresponding to the vacuum chamber. The coil radius is 6.4~cm and the linear distance between
two opposing coils is 20.4~cm. The vacuum B-field value is 1.4~Tesla at the cusps near the
center of the coils. These parameters represent a scaling of the experimental apparatus that was
tractable with available computational resources. Uniform density plasma is initially loaded in a
cube of 7.3~cm, centered in the simulation. The subsequent evolution is followed solving the full
set of Maxwell equations for the fields and classical Newton equations for the particles.
We report two cases, one with initial density equal to \SI{2e15}{cm^{-3}}, and the other with
\SI{2e14}{cm^{-3}}. An initial electron temperature is set at 135~keV and the ion temperature is set at
34~keV for both cases. We present two movies documenting the diamagnetic effect with frames
taken at every 2,500 cycles or \SI{1.27e-8}{s}. Movie 1 (M1) shows the high initial density case with
the density is shown in volume representation in colour, superimposed to a representation of
selected surfaces of constant magnitude of the magnetic field, while Movie 2 (M2) shows the
low initial density case. The rho1=1 corresponds to the plasma density of \SI{1e16}{cm^{-3}} and
magnitude of B=1 corresponds to the magnetic field of \SI{1.38e3}{T}. As time evolves, the
initial plasma expands and pushes the magnetic field out, providing clear evidence of the
diamagnetic effect. The case with lower initial density shows a weak diamagnetic response,
while the higher initial density case shows a strong response. The surface of $\beta=1$ has a particular
importance in determining the diamagnetic effect, because there the plasma and magnetic
pressure are nearly equal. This surface expands as the initial plasma loaded centrally expands
and relaxes to towards equilibrium with the magnetic field. We report such expansion in the
third movie provided in the supporting material, showing the expanding surface $\beta=1$ for the high
density case in Movie 3 (M3). Finally, we compare the precise geometry of the $\beta=1$ surface for
two cases at the cycle 10,000, corresponding to time t=\SI{5.08e-8}{s} in Movie 4 (M4) and Movie 5 (M5). The
movie rotates the view to allow the viewer to appreciate the 3D structure.

\subsection{Time resolved measurements of Tungsten impurities}
After more than 200 plasma injections, the tungsten cathode in the plasma co-axial gun
shows significant erosion, as shown in Figure S1. Since tungsten is good x-ray generating target
material when exposed to the electron beam, injection of tungsten impurities into the magnetic
cusp has been investigated in the visible spectral range from 390~nm to 450~nm with the use of
a spectrometer coupled with a gated CCD camera. Figures S2 and S3 show 8~$\mu$s time integrated
spectra from the cups plasma with the same gain setting. Figure S2 shows the plasma emission
spectrum during the initial high $\beta$ state, between 12~$\mu$s and 20~$\mu$s after the plasma injection.
Figure S3 shows the plasma emission spectrum after the plasma decay, between 42~$\mu$s and 50~$\mu$s
after the plasma injection. During initial high $\beta$ state, emission spectrum is dominated by carbon
lines with contributions from Tungsten ion lines. In comparison, emission spectrum is
dominated by tungsten neutral lines after the plasma decay. Figure S4 shows relative brightness
of carbon and tungsten lines as a function of time. The signals are normalized to their brightness
at 12~$\mu$s after the plasma injection. In addition, signals from photodiodes, filtered to look at
hydrogen and ionized carbon (H$\alpha$ and C-II), are plotted for comparison. The results indicate that
tungsten impurities are introduced to the cusp plasma region gradually and on a slower time
scale than the high $\beta$ plasma injection. This is consistent with weak x-ray emission in the
experiment with slow and gradual rises and decays.

\begin{figure}
   \includegraphics[trim=0 0 0 -25,clip,width=0.5\linewidth]{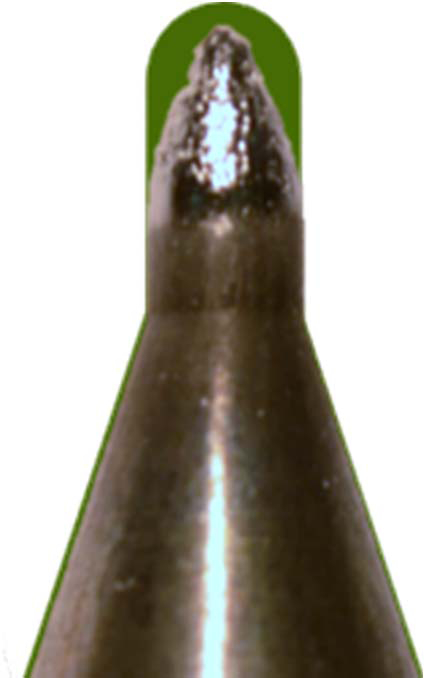}
   \caption{A close-up photograph of a tungsten cathode after about 200 plasma injections. The width of the central tip is 2~mm and the green area highlights material erosion.}
\end{figure}

\begin{figure*}
   \includegraphics[width=0.75\linewidth]{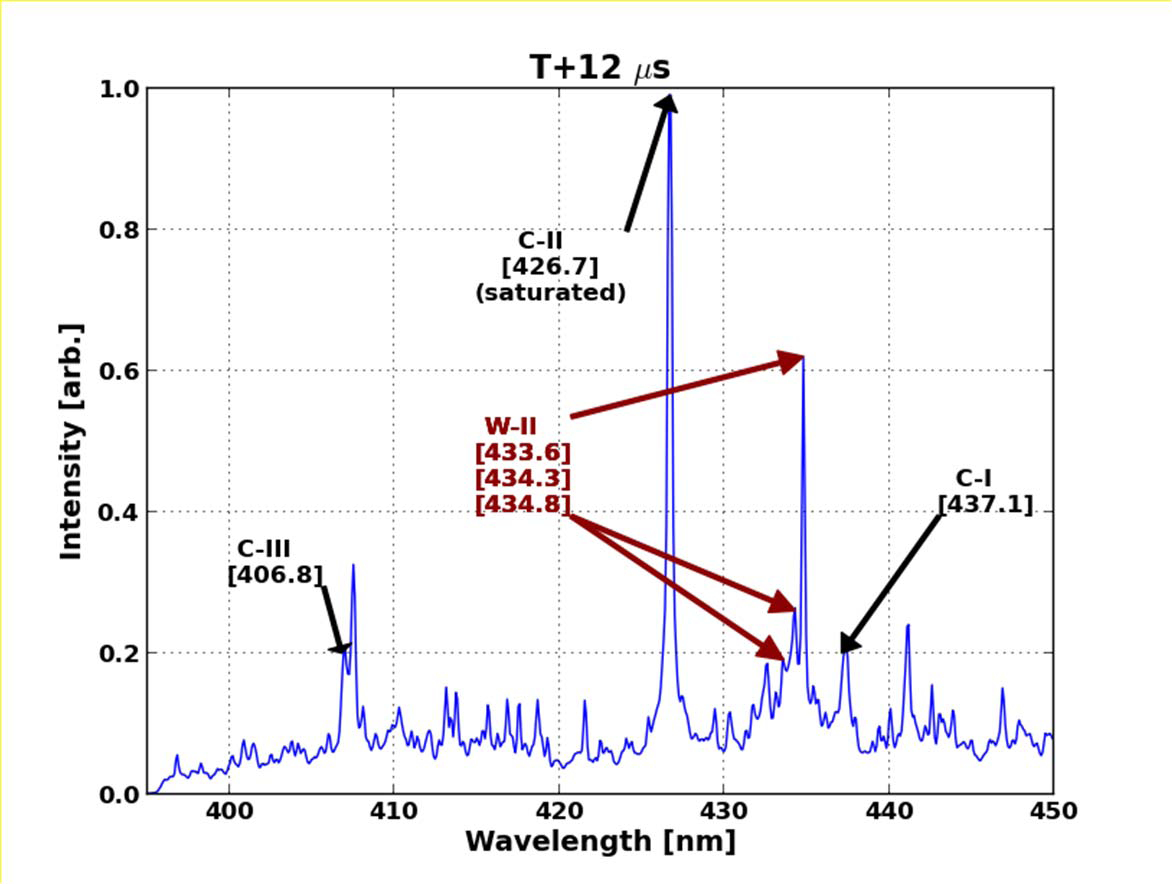}
   \caption{Time integrated spectra of the cusp plasma between 12~$\mu$s and 20~$\mu$s after the plasma injection.}
\end{figure*}

\begin{figure*}
   \includegraphics[width=0.75\linewidth]{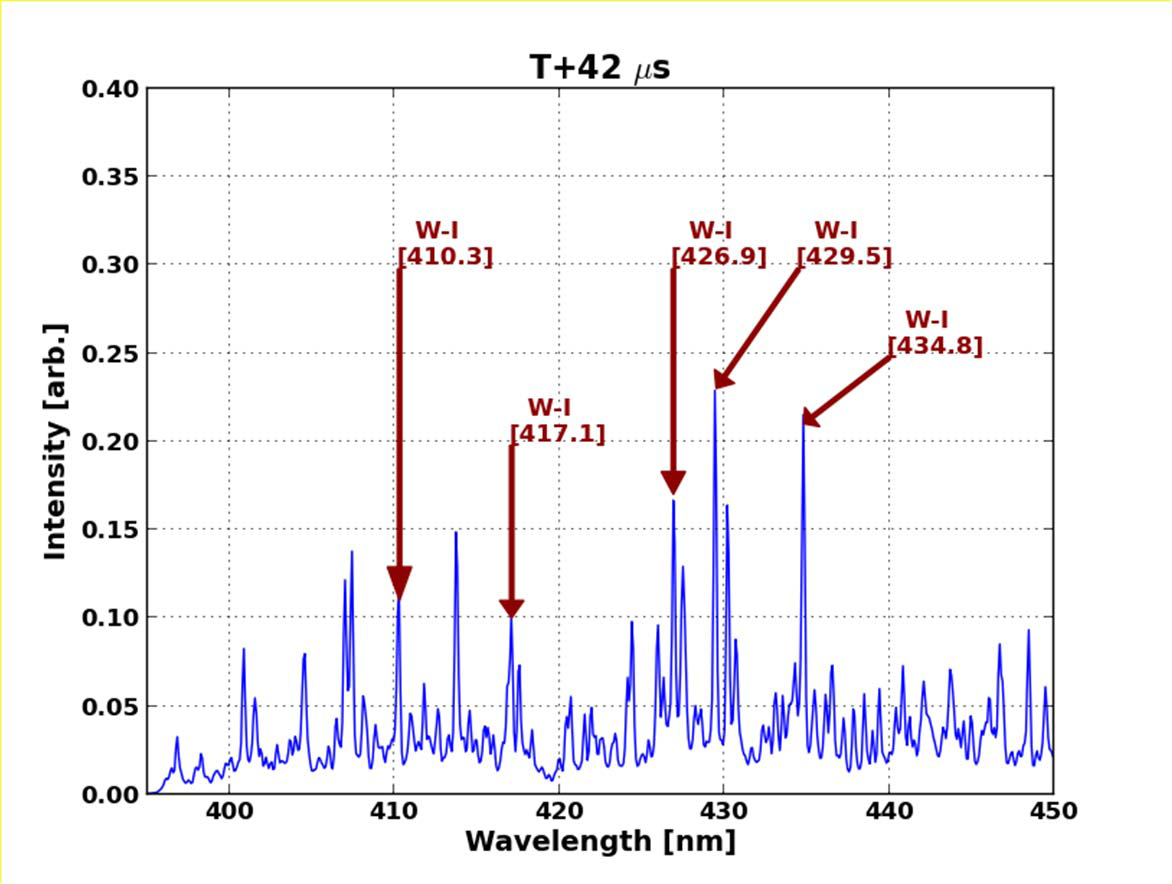}
   \caption{Time integrated spectra of the cusp plasma between 42 $\mu$s to 50 $\mu$s after the plasma injection.}
\end{figure*}

\begin{figure*}
   \includegraphics[width=0.75\linewidth]{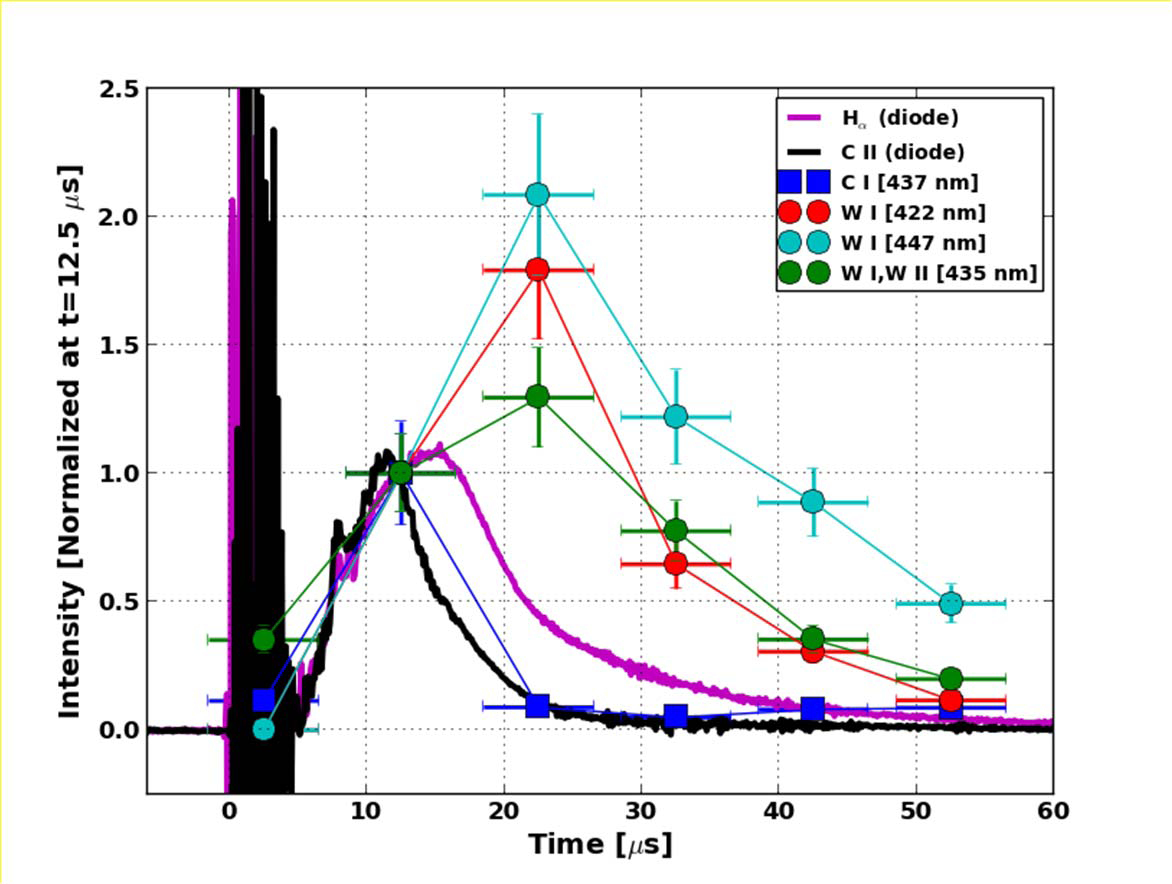}
   \caption{Temporal variation of plasma line emission intensities, from
   ensemble of six plasma shots. The magenta and black lines are the average filtered photodiode
   signals for hydrogen alpha and carbon ion lines. It is noted that the diode signals had less than a
   10\% variation from shot to shot.}
\end{figure*}

\end{document}